\documentclass[aps,twocolumn]{revtex4-2}
\usepackage{amssymb}
\usepackage{graphicx}
\usepackage{dcolumn}
\usepackage{bm}
\usepackage{amsmath}
\usepackage{soul,color}
\usepackage{textcomp}
\usepackage{times}
\usepackage[colorlinks,linkcolor=red,citecolor=blue]{hyperref}

\begin{document}

\title{Four-mode quantum sensing and Fisher information in a spin-orbit-coupled Bose gas}

\author{Fei Zhu$^{1}$}
\author{Zheng Tang$^{1}$}
\author{Liang Zeng$^{1}$}
\author{Shu Wang$^{1}$}
\author{Li Chen$^{1}$}
\email{lchen@sxu.edu.cn}

\affiliation{
$^1${Institute of Theoretical Physics, State Key Laboratory of Quantum Optics and Quantum Optics Devices, Shanxi University, Taiyuan 030006, China}\\
}

\begin{abstract}
Multi-mode squeezing and entanglement are important resources in quantum metrology and sensing. For spin-1/2 Bose-Einstein condensates subject to spin-orbit coupling (SOC), previous studies on spin squeezing have been limited to two-mode systems. 
In this work, we demonstrate that such a system can naturally construct a four-mode model spanning an $\mathfrak{su}(4)$ algebra with six SU(2) subspaces. Using spin squeezing parameters and quantum Fisher information matrices, we analyze the dynamical evolution of coherent spin states. 
The results show that, beyond two-mode models, the SOC-induced four-mode couplings give rise to richer entanglement-enhanced sensing approaching the Heisenberg limit across various SU(2) subspaces.
Additionally, by tuning a single system parameter (the Raman Rabi frequency), one can selectively control the optimal measurement directions across different subspaces.
\end{abstract}

\maketitle

\section{Introduction}
\label{sec:introduction}

Quantum squeezing and entanglement have emerged as essential resources for quantum information processing and precision metrology, enabling operations and measurements beyond classical limits \cite{degenQuantumSensing2017a,ma2011,grossSpinSqueezingEntanglement2012,horodeckiQuantumEntanglement2009}. 
Spin squeezed states (SSS), pioneered by Kitagawa and Ueda \cite{kitagawaSqueezedSpinStates1993}, are collective states of two-level atomic ensembles that can be effectively described using a two-mode model within the Schwinger bosonic representation. Spin squeezing manifests as reduced quantum fluctuation in one collective spin component at the expense of increased fluctuation in its orthogonal component, in accordance with the Heisenberg uncertainty relation.
Beyond spin squeezing, the quantum Fisher information (QFI) provides a more fundamental metric to characterize metrologically useful entanglement, providing a lower bound on the precision achievable in parameter estimations, known as the quantum Cramér-Rao bound \cite{degenQuantumSensing2017a,HELSTROM1967101,carolloQuantumnessMultiparameterQuantum2019,liuQuantumFisherInformation2020b,ALBARELLI2020126311}. 
In the recent decade, numerous important experimental advances in SSS and QFI have been reported, including entanglement-enhanced atomic clocks \cite{pedrozo-penafielEntanglementOpticalAtomicclock2020,colomboEntanglementenhancedOpticalAtomic2022, gilSpinSqueezingRydberg2014,Yang2025}, quantum magnetometers \cite{Wasilewski2010, Sewell2012, Muessel2014}, and precision interferometers that surpass the standard quantum limit \cite{hostenMeasurementNoise1002016, bornetScalableSpinSqueezing2023}.

Multi-mode systems, which involve a few (more than two) modes, generally offer richer physics compared to their two-mode counterparts. 
By extending beyond the simple two-level picture, multi-mode configurations enable the exploration of higher-dimensional Hilbert spaces and richer quantum correlations. 
In fundamental physics, multi-mode systems generally serve as tractable toy models that capture essential features of complex multi-level systems while remaining experimentally accessible. For instance, electromagnetically induced transparency \cite{fleischhauerElectromagneticallyInducedTransparency2005,Finkelstein2023} in multi-level atomic systems demonstrates quantum interference effects that cannot be observed in two-level systems, allowing for phenomena like slow light \cite{hauLightSpeedReduction1999,Bigelow2003,Naeini2011} and enhanced nonlinear optical responses \cite{Harris1990,Wang2001}. 
Quantum simulators with multiple modes have enabled the demonstration of atomic transport \cite{Caliga2016}, quantum random walk \cite{Peruzzo2011, Defienne2016}, and the properties of lattice gauge fields \cite{Martinez2016}.
In quantum information processing, multi-mode entanglement structures are valuable for quantum sensing capabilities, multi-parameter estimation, and novel quantum information protocols \cite{kokRoleEntanglementCalibrating2017, wilsonOneaxisTwistingSimultaneous2022, pezzeOptimalMeasurementsSimultaneous2017, 
reillyOptimalGeneratorsQuantum2023}. 
For example, W states have been demonstrated to achieve multi-parameter estimation with sensitivity approaching the Heisenberg limit \cite{hongQuantumEnhancedMultiplephase2021}.

The realization of spin-orbit coupling (SOC) in cold atomic gases represents a significant advancement in cold atomic physics over the past decade \cite{Galitski2013,Goldman2014,Zhai2015,WZhang2018,Chen2023}, which enables precise control of atomic spin and orbital degrees of freedom, as well as their interactions, through two-photon Raman processes \cite{linSpinOrbitCoupled2011}, leading to non-trivial energy-band structures.
Working along with inter-atomic interactions, such systems have enabled remarkable observations, including the realization of non-Abelian gauge fields and exotic quantum phases of matter such as the striped supersolid-like phases \cite{Ho2011, Li2012, Zheng2013, Li2017}, topological vortices and various topological phase transitions \cite{Peng2022, DeMarco2015, Sun2015, Chen2016, Chen2020SN, KJChen2020, HChen2018, PChen2018, Zhang2018}. 
When the Bose-Einstein condensates (BEC) is restricted to the two lowest momentum modes, it effectively forms a typical one-axis twisting (OAT) model exhibiting spin squeezing \cite{chenSpinSqueezingSpinorbitcoupled2020}. 
However, the potential of SOC BECs for creating and manipulating quantum entanglement across multiple modes remains unexplored.

In this work, we demonstrate that the spin-1/2 SOC BEC naturally supports a four-mode model by incorporating the two additional momentum states in the upper energy band.
This configuration forms an $\mathfrak{su}(4)$ algebraic structure containing six SU(2) subspaces respectively labeled by 
$\{\boldsymbol{\mathcal{K}},\boldsymbol{\mathcal{P}},\boldsymbol{\mathcal{E}},\boldsymbol{\mathcal{M}},\boldsymbol{\mathcal{J}},\boldsymbol{\mathcal{N}}\}$
, dramatically expanding the available quantum correlations compared to the conventional two-mode system with only one SU(2) space $\boldsymbol{\mathcal{K}}$. We analyze the dynamical evolution of coherent spin states initialized in different subspaces, and characterize the state properties through both spin squeezing parameters and QFI spectra. 
The results show that the SOC-induced four-mode coupling enables quantum metrological advantage to be realized in all SU(2) subspaces, even when the initial state has particle occupations in only two modes. 
Furthermore, by tuning the SOC parameter (i.e., the Raman Rabi frequency), the optimal measurement directions can traverse across different subspaces, with the maximum QFI spectrum approaching the Heisenberg scaling limit. 
Our findings demonstrate that the SOC can serve as an important control knob for spinor BECs, facilitating quantum metrology and sensing in specific subspaces.

The rest of the paper is organized as follows. In Sec.~\ref{sec:theory}, we establish the theoretical framework by introducing the four-mode SOC BEC model, deriving the effective Hamiltonian, and constructing the $\mathfrak{su}(4)$ algebraic structure with six SU(2) subspaces. We also define the quantum metrological quantities including spin squeezing parameters and quantum Fisher information matrices. Sec.~\ref{sec:results} presents our main results, analyzing the dynamics of coherent spin states in both $\boldsymbol{\mathcal{K}}$-space and $\boldsymbol{\mathcal{E}}$-space, examining spin squeezing evolution, QFI spectra, and optimal measurement directions across different subspaces. Finally, Sec.~\ref{sec:conclusion} discusses experimental considerations and concludes with a summary of our key findings.

\section{Theoretical Framework}
\label{sec:theory}

\subsection{Model}
\label{sec:model}

We consider a quasi-one-dimensional spin-1/2 BEC with spin-orbit coupling realized by a pair of counter-propagating Raman lasers along the $x$-axis. In the laboratory frame, the single-particle Hamiltonian along the $x$-direction is in the well-known form (setting $\hbar = m = 1$)
\begin{equation}
    h_0 = \frac{\left(k-k_r \sigma_z\right)^2}{2}+\frac{\Omega}{2} \sigma_x,
    \label{eq:single_particle_hamiltonian}
\end{equation}
where $k$ is the momentum of atoms along the SOC direction, $k_r$ is the recoil momentum of the Raman lasers, $\Omega$ is the Raman Rabi frequency, and $\sigma_i$ ($i=x,y,z$) are the Pauli matrices acting on the two internal pseudo-spin states. The momentum $k$ is conserved for $h_0$, i.e. $\left[k,h_0\right] = 0$. 

\begin{figure}[t]
    \centering
    \includegraphics[width=\linewidth]{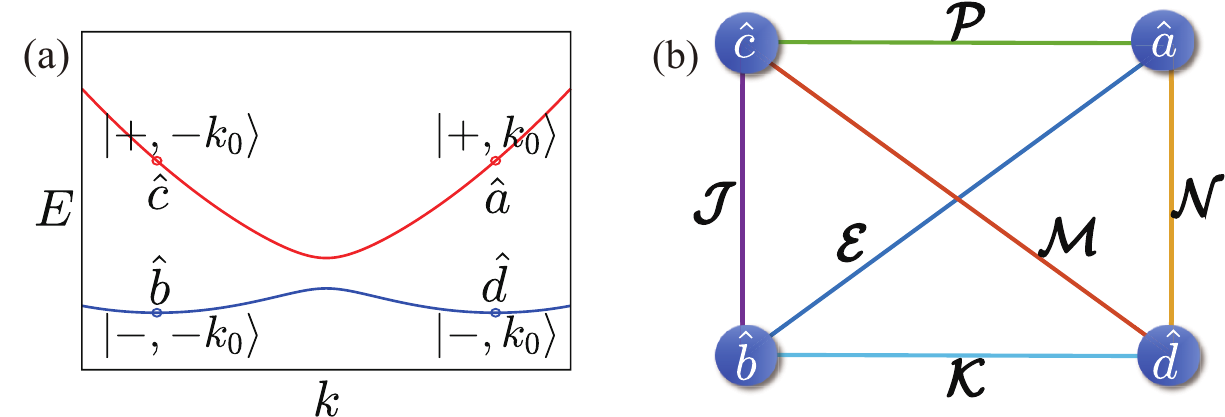}
    \caption{(a) Energy band structure of a SOC BEC system showing the four modes with circles. The operators $\hat{a}$, $\hat{b}$, $\hat{c}$, and $\hat{d}$ correspond to the four modes $|+,k_0\rangle$, $|-,-k_0\rangle$, $|+,-k_0\rangle$, and $|-,k_0\rangle$, respectively. 
    (b) The six SU(2) subspaces of the four-mode system: $\boldsymbol{\mathcal{K}}$ (constructed by $\hat{b}$ and $\hat{d}$ modes), $\boldsymbol{\mathcal{P}}$ ($\hat{c}$ and $\hat{a}$ modes), $\boldsymbol{\mathcal{E}}$ ($\hat{a}$ and $\hat{b}$ modes), $\boldsymbol{\mathcal{M}}$ ($\hat{c}$ and $\hat{d}$ modes), $\boldsymbol{\mathcal{J}}$ ($\hat{b}$ and $\hat{c}$ modes), and $\boldsymbol{\mathcal{N}}$ ($\hat{a}$ and $\hat{d}$ modes).
    }
    \label{fig:four_mode_system}
\end{figure}

The single-particle spectrum is in two bands as illustrated in Fig.~\ref{fig:four_mode_system}(a), with $E_{\pm}(k) = \frac{k^2}{2}+E_r\left(1 \pm \sqrt{\frac{2 k^2}{E_r}+\frac{\Omega^2}{4 E_r^2}}\right)$, where $E_r = k_r^2/2$ is the recoil energy. 
For $0 < \Omega < 4E_r$, the lower energy band $E_{-}(k)$ exhibits a double-well structure, with two degenerate minima located at $\pm k_0$ with $k_0 = k_r \sqrt{1 - \Omega^2/(16 E_r^2)}$. These two states are respectively labeled as $|-,\pm k_0\rangle$, which form the basis of the commonly studied two-mode model in previous works. 
For $\Omega \geq 4E_r$, $E_{-}(k)$ transitions from a double-well to a single-well structure with a global minimum at $k_0=0$. 
The change of spectral structure would eventually lead to a quantum phase transition from the plane-wave phase to the zero-momentum phase when two-body atomic interactions are included, which has been extensively discussed in Refs.~\cite{Li2012}.

We extend the conventional two-mode model to include the corresponding states in the upper band $E_+$, namely $|+,\pm k_0\rangle$, thus forming a four-mode model, as labeled by circles in Fig.~\ref{fig:four_mode_system}(a).
This is a natural extension, since the nonlinear dispersion of both upper and lower energy bands ensures that, for initial states with momenta $\pm k_0$, momentum and energy conservations make transitions to other off-resonant momentum states less likely.
The four momentum modes can be expressed in the bare spin-momentum basis as
\begin{equation}
    \begin{aligned}
    & \left|+, k_0\right\rangle=\binom{\sin \theta}{\cos \theta}\left|k_0\right\rangle, \ \ 
    \left|-,-k_0\right\rangle=\binom{\sin \theta}{-\cos \theta}\left|-k_0\right\rangle, \\
    & \left|+,-k_0\right\rangle=\binom{\cos \theta}{\sin \theta}\left|-k_0\right\rangle,\ \
    \left|-, k_0\right\rangle=\binom{\cos \theta}{-\sin \theta}\left|k_0\right\rangle,
    \end{aligned}
    \label{eq:four_mode_basis}
\end{equation}
where $\theta = \arccos(k_0/k_r)/2$ is the spin-momentum dressing angle determined by the Raman coupling strength. Note that, the angle $\theta$ varies from 0 to $\pi/4$ as $\Omega$ increases from 0 to $4E_r$. In the real-space, the wave function of the momentum state is the plane wave, i.e., $\langle x|k\rangle = e^{ikx}/\sqrt{L}$, with $L$ being the system size.

For an interacting BEC with two-body s-wave interactions, the many-body Hamiltonian can be written as
\begin{equation}
    H = H_0 + H_1,
    \label{eq:many_body_hamiltonian}
\end{equation}
where $H_0 = \int  dx \hat{\boldsymbol{\Psi}}^{\dagger}(x) h_0 \hat{\boldsymbol{\Psi}}(x)$ with $\hat{\boldsymbol{\Psi}}(x) = \{\hat{\psi}_{\uparrow}(x), \hat{\psi}_{\downarrow}(x)\}^T$ being the spinor field operator with two pseudo-spin components $\uparrow$ and $\downarrow$, and 
\begin{equation}
    H_1=\int dx \left[\frac{g}{2} \left(\hat{\psi}_{\uparrow}^{\dagger} \hat{\psi}_{\uparrow}^{\dagger} \hat{\psi}_{\uparrow} \hat{\psi}_{\uparrow}+\hat{\psi}_{\downarrow}^{\dagger} \hat{\psi}_{\downarrow}^{\dagger} \hat{\psi}_{\downarrow} \hat{\psi}_{\downarrow}\right)+g_{\uparrow\downarrow} \hat{\psi}_{\uparrow}^{\dagger} \hat{\psi}_{\downarrow}^{\dagger} \hat{\psi}_{\uparrow} \hat{\psi}_{\downarrow}\right]
    \label{eq:interaction_hamiltonian}
\end{equation}
is the interaction Hamiltonian, with $g$ and $g_{\uparrow\downarrow}$ being the intra- and inter-spin interaction strengths, respectively. 
Here, we assume $g_{\uparrow\downarrow} = g$, which is a good approximation for the most commonly used alkaline-metal atoms in experiments, such as $^{87}$Rb or $^{23}$Na where $g$ and $g_{\uparrow\downarrow}$ differ by less than $1\%$ \cite{Kawaguchi2012,linSpinOrbitCoupled2011}.
Next, we introduce the bosonic operators $\hat{a}$, $\hat{b}$, $\hat{c}$, and $\hat{d}$ respectively corresponding to the four dressed modes, i.e., $|+,k_0\rangle$, $|-,-k_0\rangle$, $|-,k_0\rangle$, and $|+,-k_0\rangle$, which have been defined in Eq.~(\ref{eq:four_mode_basis}) and illustrated in Fig.~\ref{fig:four_mode_system}. Then, the field operator can be approximately expanded in the four-mode basis as
\begin{equation}
    \begin{aligned}
    \hat{\boldsymbol{\Psi}} \approx &\binom{\sin \theta}{\cos \theta}\left|k_0\right\rangle \hat{a}+\binom{\sin \theta}{-\cos \theta}\left|-k_0\right\rangle \hat{b} \\
    & +\binom{\cos \theta}{\sin \theta}\left|-k_0\right\rangle \hat{c}+\binom{\cos \theta}{-\sin \theta}\left|k_0\right\rangle \hat{d}
    \end{aligned}.
    \label{eq:field_operator_expansion}
\end{equation}
In such a framework, the single-particle Hamiltonian is in a diagonal form
\begin{equation}
    H_0 = E_+(k_0) (\hat{a}^{\dagger}\hat{a} + \hat{c}^{\dagger}\hat{c}) + E_-(k_0) (\hat{b}^{\dagger}\hat{b} + \hat{d}^{\dagger}\hat{d}),
    \label{eq:single_particle_hamiltonian_diagonal}
\end{equation}
and the interaction Hamiltonian $H_1$ (after some tedious but straightforward calculations) is given by
\begin{equation}
    \begin{aligned}
    H_1 &= \chi\left[\cos (2 \theta)\left(\hat{a}^{\dagger} \hat{b}-\hat{d}^{\dagger} \hat{c}\right)-\sin (2 \theta)\left(\hat{a}^{\dagger} \hat{c}+\hat{d}^{\dagger} \hat{b}\right)\right] \\
    &\times \left[\cos (2 \theta)\left(\hat{b}^{\dagger} \hat{a}-\hat{c}^{\dagger} \hat{d}\right)-\sin (2 \theta)\left(\hat{c}^{\dagger} \hat{a}+\hat{b}^{\dagger} \hat{d}\right)\right]
    \end{aligned}
    \label{eq:interaction_hamiltonian_four_mode}
\end{equation}
up to some globally conserved quantities, such as the total particle number $\hat{N} = \sum_{\alpha}\hat{N}_\alpha$ and the total momentum $\hat{N}_a + \hat{N}_c - \hat{N}_b - \hat{N}_d$, which have no effect on the dynamics. 
Here, $\hat{N}_{\alpha} = \hat{\alpha}^{\dagger}\hat{\alpha}$ is the particle number operator for the mode $\alpha \in \{a,b,c,d\}$ and $\chi = g/L$.

We eventually enter the rotating frame via a unitary transformation $U(t)=\exp\{i [E_+(k_0)- E_-(k_0)] t (\hat{N}_a+ \hat{N}_c - \hat{N}_b - \hat{N}_d )/2\}$ which eliminate $H_0$ and transforms the Hamiltonian to
\begin{equation}
    \begin{aligned}
    H_\text{eff} &= \chi \left[\cos ^2(2 \theta)\left(\hat{N}_a \hat{N}_b+\hat{N}_d\hat{N}_c\right)\right. \\
    &\left. + \sin ^2(2 \theta)\left(\hat{N}_a \hat{N}_c+\hat{N}_b \hat{N}_d+\hat{a}^{\dagger} \hat{b}^{\dagger} \hat{c} \hat{d}+\hat{d}^{\dagger} \hat{c}^{\dagger} \hat{b} \hat{a}\right)\right].
    \end{aligned}
    \label{eq:hamiltonian_interaction_picture}
\end{equation} 
It is evident that Eq.~(\ref{eq:hamiltonian_interaction_picture}) contains both two-mode scattering (involving two particle number operators) and four-mode scattering (the last two terms). 
The four-mode scattering terms will play a crucial role in inducing quantum sensing capabilities across multiple SU(2) subspaces as we will show in the following sections.
As expected, the total momentum is always conserved, manifested as the total momentum remaining unchanged before the scattering (particle pair annihilation) and after the scattering (particle pair creation).
One can additionally check that, in the absence of the two upper-band modes, the four-mode Hamiltonian in Eq.~(\ref{eq:hamiltonian_interaction_picture}) can be readily reduced to the two-mode one-axis twisting Hamiltonian $\propto -\chi \sin^2 (2\theta)(\hat{N}_b-\hat{N}_d)^2$ of the SOC BEC which has been studied in Ref.~\cite{chenSpinSqueezingSpinorbitcoupled2020}.

\subsection{Algebraic Structure}
\label{sec:algebra}

In constrast to conventional two-mode modes living in a single SU(2) space, the dynamics of a four-mode system is generically characterized by the SU(4) Lie group generated by 15 orthogonal generators labeled by $\hat{G}_{i=1,\cdots,15}$. The commutators (also called the structure factors) of these generators constitute the algebraic structure. The $\mathfrak{su}(4)$ algebra has 3 mutually commuting Cartan generators, typically denoted as $\mathbf{C} = \{\hat{C}_1, \hat{C}_2, \hat{C}_3\}$, and 6 pairs of opposite roots, typically denoted as $\hat{R}_{\pm\boldsymbol{\lambda}}$, where $\boldsymbol{\lambda}$ are three-dimensional vectors. 
Each pair of roots together with the Cartan generators spans a closed SU(2) subspace, satisfying $[\hat{C}_i, \hat{R}_{\pm\boldsymbol{\lambda}}] = \lambda_i \hat{R}_{\pm\boldsymbol{\lambda}}$ and $[\hat{R}_{\boldsymbol{\lambda}}, \hat{R}_{-\boldsymbol{\lambda}}] = \boldsymbol{\lambda} \cdot \mathbf{C}$, reminiscent the traditional relations $[\hat{S}_z, \hat{S}_{\pm}] =\pm \hat{S}_{\pm}$ and $[\hat{S}_+, \hat{S}_-] = 2 \hat{S}_z$ of SU(2) spin-angular-momentum operators.

In the current four-mode system, we can construct subspaces using $6$ mode pairs, as illustrated in Fig.~\ref{fig:four_mode_system}(b), where each line connecting two nodes represents one subspace. The 6 subspaces are specifically labeled by: 
$\boldsymbol{\mathcal{K}}$, $\boldsymbol{\mathcal{P}}$, $\boldsymbol{\mathcal{E}}$, $\boldsymbol{\mathcal{M}}$, $\boldsymbol{\mathcal{J}}$, and $\boldsymbol{\mathcal{N}}$. 
Then we have the 12 roots  corresponding to the raising and lowering operators defined as the coupling between the connected-modes. For example, the raising and lowering operators of $\boldsymbol{\mathcal{K}}$ subspace are $\hat{\mathcal{K}}_+ = \hat{d}^{\dagger}\hat{b}$ and $\hat{\mathcal{K}}_- = \hat{b}^{\dagger}\hat{d}$, based on which one can construct $\hat{\mathcal{K}}_x = \frac{1}{2}(\hat{\mathcal{K}}_+ + \hat{\mathcal{K}}_-)$, $\hat{\mathcal{K}}_y = \frac{-i}{2}(\hat{\mathcal{K}}_+ - \hat{\mathcal{K}}_-)$, and $\hat{\mathcal{K}}_z = \frac{1}{2}(\hat{d}^{\dagger}\hat{d} - \hat{b}^{\dagger}\hat{b})$, which obviously satisfy the SU(2) commutating relations.
Furthermore, we can also find the 3 Cartan generators $\hat{\mathcal{E}}_z = (\hat{a}^{\dagger}\hat{a} - \hat{b}^{\dagger}\hat{b})/2$, $\hat{\mathcal{M}}_z = (\hat{c}^{\dagger}\hat{c} - \hat{d}^{\dagger}\hat{d})/2$, and $\hat{\mathcal{U}}_z = (\hat{a}^{\dagger}\hat{a} + \hat{b}^{\dagger}\hat{b} - \hat{c}^{\dagger}\hat{c} - \hat{d}^{\dagger}\hat{d})/2\sqrt{2}$.
Eventually, we have the detailed definition of all the 15 generators of $\mathfrak{su}(4)$ algebra as listed in Table.~\ref{tab:SU4_operators}.
One can easily verify that the commutators of these generators reproduce the conventional $\mathfrak{su}(4)$ structure factors.

Note that, for conventional two-mode models defined on the lower energy band \cite{chenSpinSqueezingSpinorbitcoupled2020}, only the $\boldsymbol{\mathcal{K}}$ space exists. In contrast, our four-mode system exhibits six distinct SU(2) subspaces ready for metrology and sensing. More generally, for a $n$-mode system, there are $n(n-1)/2$ SU(2) subspaces available, as the $\mathfrak{su}(n)$ algebra contains precisely $n(n-1)/2$ pairs of roots.

\begin{table}[t]
\centering
\caption{Generators of the $\mathfrak{su}(4)$ algebra expressed in terms of bosonic operators for the four modes.}
\label{tab:SU4_operators}
\begin{tabular}{|c|c|c|}
\hline
Generator No. & Generator & Schwinger Boson Expression \\
\hline
1 & $\hat{\mathcal{K}}_x$ & $\frac{1}{2}(\hat{d}^{\dagger}\hat{b} + \hat{b}^{\dagger}\hat{d})$ \\
2 & $\hat{\mathcal{K}}_y$ & $\frac{-i}{2}(\hat{d}^{\dagger}\hat{b} - \hat{b}^{\dagger}\hat{d})$ \\
\hline
3 & $\hat{\mathcal{P}}_x$ & $\frac{1}{2}(\hat{a}^{\dagger}\hat{c} + \hat{c}^{\dagger}\hat{a})$ \\
4 & $\hat{\mathcal{P}}_y$ & $\frac{-i}{2}(\hat{a}^{\dagger}\hat{c} - \hat{c}^{\dagger}\hat{a})$ \\
\hline
5 & $\hat{\mathcal{E}}_x$ & $\frac{1}{2}(\hat{a}^{\dagger}\hat{b} + \hat{b}^{\dagger}\hat{a})$ \\
6 & $\hat{\mathcal{E}}_y$ & $\frac{-i}{2}(\hat{a}^{\dagger}\hat{b} - \hat{b}^{\dagger}\hat{a})$ \\
7 & $\hat{\mathcal{E}}_z$ & $\frac{1}{2}(\hat{a}^{\dagger}\hat{a} - \hat{b}^{\dagger}\hat{b})$ \\
\hline
8 & $\hat{\mathcal{M}}_x$ & $\frac{1}{2}(\hat{c}^{\dagger}\hat{d} + \hat{d}^{\dagger}\hat{c})$ \\
9 & $\hat{\mathcal{M}}_y$ & $\frac{-i}{2}(\hat{c}^{\dagger}\hat{d} - \hat{d}^{\dagger}\hat{c})$ \\
10 & $\hat{\mathcal{M}}_z$ & $\frac{1}{2}(\hat{c}^{\dagger}\hat{c} - \hat{d}^{\dagger}\hat{d})$ \\
\hline
11 & $\hat{\mathcal{J}}_x$ & $\frac{1}{2}(\hat{c}^{\dagger}\hat{b} + \hat{b}^{\dagger}\hat{c})$ \\
12 & $\hat{\mathcal{J}}_y$ & $\frac{-i}{2}(\hat{c}^{\dagger}\hat{b} - \hat{b}^{\dagger}\hat{c})$ \\
\hline
13 & $\hat{\mathcal{N}}_x$ & $\frac{1}{2}(\hat{a}^{\dagger}\hat{d} + \hat{d}^{\dagger}\hat{a})$ \\
14 & $\hat{\mathcal{N}}_y$ & $\frac{-i}{2}(\hat{a}^{\dagger}\hat{d} - \hat{d}^{\dagger}\hat{a})$ \\
\hline
15 & $\hat{\mathcal{U}}_z$ & $\frac{1}{2\sqrt{2}}(\hat{a}^{\dagger}\hat{a} + \hat{b}^{\dagger}\hat{b} - \hat{c}^{\dagger}\hat{c} - \hat{d}^{\dagger}\hat{d})$ \\
\hline
\end{tabular}
\end{table}

\subsection{Quantum Metrological Quantities}
\label{sec:metrics}

To characterize the metrological utility of the four-mode system, we employ two metrics: spin squeezing parameters and quantum Fisher information matrices, with the former being applicable to a SU(2) subspace, whereas the latter being applicable to the both SU(2) and SU(4) spaces.

For a certain SU(2) subspace $\boldsymbol{\mathcal{S}} \in \{\boldsymbol{\mathcal{K}}, \boldsymbol{\mathcal{P}}, \boldsymbol{\mathcal{E}}, \boldsymbol{\mathcal{M}}, \boldsymbol{\mathcal{J}}, \boldsymbol{\mathcal{N}}\}$, the Kitagawa-Ueda spin squeezing parameter \cite{kitagawaSqueezedSpinStates1993} is defined as
\begin{equation}
    \xi_{S,\mathcal{S}}^2 = \frac{4 \min\left( \Delta \hat{\mathcal{S}}_\perp^2 \right)}{N}
    \label{eq:kitagawa_ueda_parameter}
\end{equation}
where the minimization is over all directions on the Bloch sphere perpendicular to the mean spin direction $\boldsymbol{s} = \langle \hat{\boldsymbol{\mathcal{S}}} \rangle / |\langle \hat{\boldsymbol{\mathcal{S}}} \rangle|$, and $N/4$ being the quantum variance of coherent spin states \cite{ma2011}. 
On the other hand, the Wineland spin squeezing parameter is defined as \cite{winelandSpinSqueezingReduced1992,winelandSqueezedAtomicStates1994}
\begin{equation}
   \xi_{R,\mathcal{S}}^2 = \frac{N \min \left( \Delta \hat{\mathcal{S}}_\perp^2 \right)}{|\langle \hat{\mathcal{S}} \rangle|^2}.
   \label{eq:wineland_parameter}
\end{equation}
The condition $\xi_{S,\mathcal{S}}^2 < 1$ indicates the presence of spin squeezing according to the variance-based criterion, while $\xi_{R,\mathcal{S}}^2 < 1$ signifies metrological spin squeezing that provides quantum-enhanced sensitivity beyond the standard quantum limit $\propto 1/\sqrt{N}$. Acoording to definition, one generally has $\xi_{S,\mathcal{S}}^2 \le \xi_{R,\mathcal{S}}^2$.

The quantum Fisher information provides another characterization of metrological utility through the quantum Cramér-Rao bound \cite{degenQuantumSensing2017a, ma2011}. For parameter estimation with encoding transformation $e^{-i \phi \hat{G}}$, where $\phi$ is the parameter to be measured and $\hat{G}$ is a certain generator operator, the ultimate precision limit is governed by
$\Delta \phi \geq \nu^{-1/2} {F}_Q^{-1/2}$,
where $\nu$ is the number of measurements and ${F}_Q$ is the QFI. For pure states, the QFI matrix is defined as
\begin{equation}
    [\mathbf{F}_Q]_{ij} = 4 \left[ \langle \hat{G}_i \hat{G}_j \rangle - \langle \hat{G}_i \rangle \langle \hat{G}_j \rangle \right],
    \label{eq:qfi_matrix_elements}
\end{equation}
whose eigenvalue spectrum determines the metrological capabilities: eigenvalues $\lambda_i > N$ indicate quantum sensing capabilities, while those $\lambda_i \to N^2$ signal optimal quantum enhancement at the Heisenberg scaling limit \cite{pezzeEntanglementNonlinearDynamics2009,degenQuantumSensing2017a}. 
The corresponding eigenvectors $\mathbf{v}_i$ identify optimal measurement directions \cite{reillyOptimalGeneratorsQuantum2023}, with the optimal generator given by $\hat{G}_{i}^{\text{opt}} = \mathbf{v}_i \cdot \hat{\mathbf{G}}$. In our following, we calculate QFI at two levels: 1) A comprehensive $15 \times 15$ matrix $\mathbf{F}_Q$ using all SU(4) generators listed in Table~\ref{tab:SU4_operators}, with the corresponding eigenspectra and eigenvectors being $\lambda_{i = \{1,\cdots,15\}}$ and $\mathbf{v}_{i}$; 2) $3 \times 3$ QFI matrices $\mathbf{F}_Q^{({\mathcal{S}})}$ for each SU(2) subspace using generators $\{\hat{{\mathcal{S}}}_x, \hat{{\mathcal{S}}}_y, \hat{{\mathcal{S}}}_z\}$, whose eigenspectra and eigenvectors are respectively denoted as $\lambda_{i = \{1,2,3\}}^{(\mathcal{S})}$ and $\mathbf{v}_{i}^{(\mathcal{S})}$;

\section{Results}
\label{sec:results}

Based on the Hamiltonian [Eq.~(\ref{eq:hamiltonian_interaction_picture})] and the theoretical framework established above, we calculate the dynamics of the four-mode system. Numerically, we expand both the Hamiltonian and the initial state $|\psi(0)\rangle$ in the Fock basis $|n_a, n_b, n_c, n_d\rangle$ and solve the Schrödinger equation through exact diagonalization, where $n_\alpha$ represents the quantum number of the particle number operator $\hat{N}_\alpha$ and satisfies the particle number conservation condition $\sum_\alpha n_\alpha = N$. We will examine the dynamics of spin squeezing and QFI spectra, as well as their dependence on the dressing parameter $\theta$.

Since momentum $k$ is conserved, initial states with definite momentum, such as the plane wave state $\propto (\hat{b}^{\dagger})^N |0\rangle$, would exhibit trivial dynamics. In the following discussions, we focus on two types initial states: the $\boldsymbol{\mathcal{K}}$-space coherent spin state (CSS) defined as
\begin{equation}
    |\text{CSS}\rangle_{\boldsymbol{\mathcal{K}}} = \frac{1}{\sqrt{N!}} \left(\frac{\hat{b}^{\dagger} + \hat{d}^{\dagger}}{\sqrt{2}}\right)^N |\mathbf{0}\rangle,
    \label{eq:CSS_K}
\end{equation}
and the $\boldsymbol{\mathcal{E}}$-space CSS defined as
\begin{equation}
    |\text{CSS}\rangle_{\boldsymbol{\mathcal{E}}} = \frac{1}{\sqrt{N!}} \left(\frac{\hat{a}^{\dagger} + \hat{b}^{\dagger}}{\sqrt{2}}\right)^N |\mathbf{0}\rangle.
    \label{eq:CSS_E}
\end{equation}
Both represent an equal-weight superposition of modes with the opposite momentum $\pm k_0$. For $\theta \neq 0$ (i.e., $\Omega \neq 0$), according to Eq.~(\ref{eq:four_mode_basis}), $|\text{CSS}\rangle_{\boldsymbol{\mathcal{K}}}$ state corresponds to the traditional stripe state \cite{Li2012,Zheng2013} where the up and down spin components have balanced particle occupation, whereas the state $|\text{CSS}\rangle_{\boldsymbol{\mathcal{E}}}$ denotes the unbalanced stripe state with up- and down-occupations being $N \sin^2(\theta)$ and $ N \cos^2(\theta)$, respectively.

\begin{figure}[t]
    \centering
    \includegraphics[width=\linewidth]{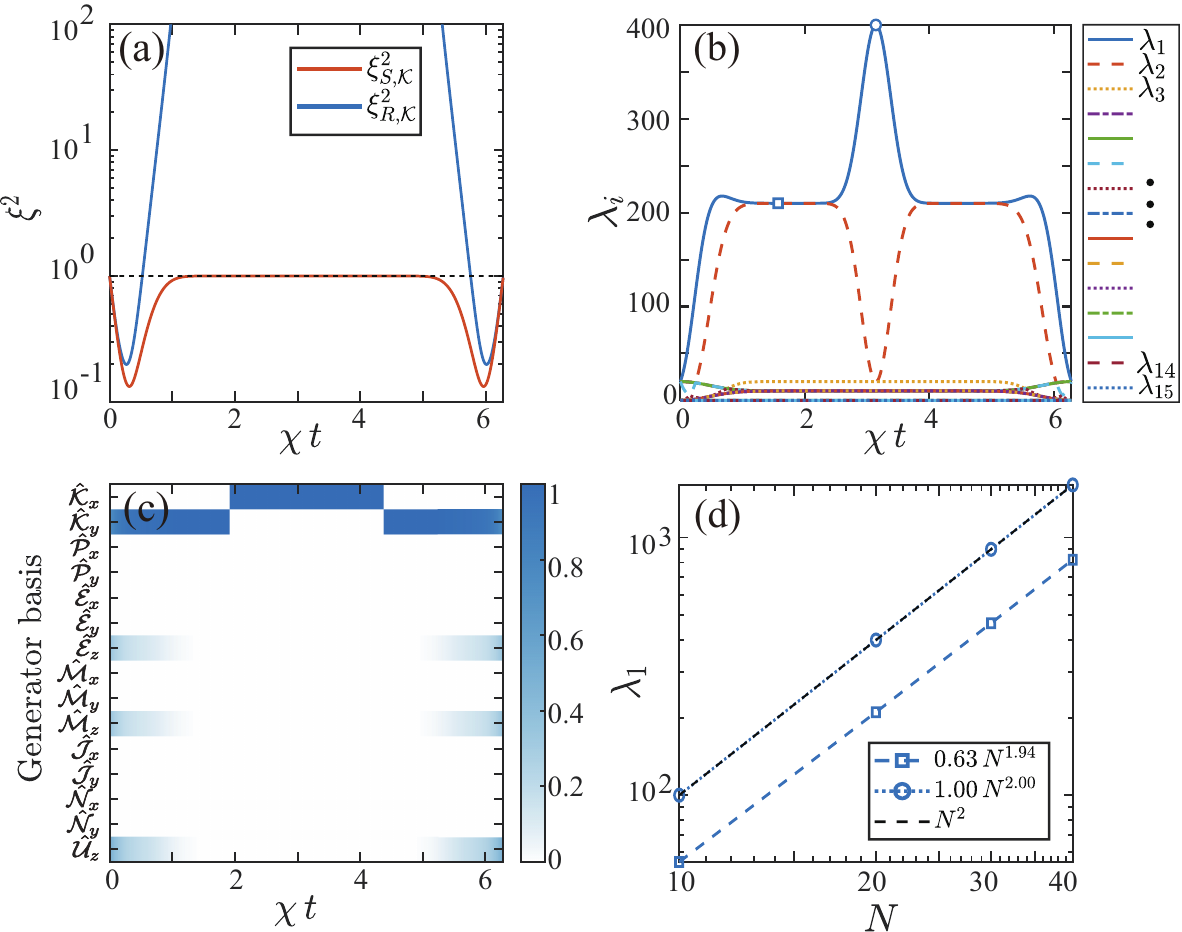}
    \caption{Dynamics of $\boldsymbol{\mathcal{K}}$-space coherent spin state for $\theta = \pi/8$ and $N = 20$. (a) Dynamics of the Kitagawa-Ueda squeezing parameter $\xi_{S,\mathcal{K}}^2$ (red line) and Wineland squeezing parameter $\xi_{R,\mathcal{K}}^2$ (blue line). The dashed horizontal line indicates the standard quantum limit. (b) QFI eigenvalue spectrum evolution with eigenvalues sorted in descending order ($\lambda_1 > \lambda_2 > \cdots > \lambda_{15}$). (c) Evolution of the projection coefficients of first eigenvector $\mathbf{v}_1$ onto the 15 generator basis. The colorbar depth represents the amplitude of projection coefficients. (d) Scaling behavior of the maximum eigenvalue with particle number $N$ at  $\ t = T/4 $  (square markers) and $\ t = T/2 $ (hollow circles) .}
    \label{fig:K_dynamics}
\end{figure}

\subsection{Dynamics of $\boldsymbol{\mathcal{K}}$-Space Coherent Spin State}
\label{sec:k_dynamics}

We first examine the dynamics starting from the $\boldsymbol{\mathcal{K}}$-space coherent spin state $|\text{CSS}\rangle_{\boldsymbol{\mathcal{K}}}$ defined in Eq.~(\ref{eq:CSS_K}), with fixed total particle number $N = 20$. For $\theta = 0$, all SU(2) subspaces exhibit neither spin squeezing nor quantum-enhanced sensitivity.
When $\theta$ is turned on, periodic spin squeezing emerges in the $\boldsymbol{\mathcal{K}}$ subspace with period $T = \pi/(\chi \sin^2(2\theta))$, resembling one-axis twisting dynamics. For $\theta = \pi/8$ (with corresponding $T = 2\pi/\chi$), Fig.~\ref{fig:K_dynamics}(a) shows the temporal evolution of both the Kitagawa-Ueda squeezing parameter $\xi_{S,\mathcal{K}}^2$ (red line) and the Wineland squeezing parameter $\xi_{R,\mathcal{K}}^2$ (blue line). The system achieves significant spin squeezing with $\{\xi_{S,\mathcal{K}}^2,\xi_{R,\mathcal{K}}^2\} < 1$ and $\xi_{R,\mathcal{K}}^2 \geq \xi_{S,\mathcal{K}}^2$, as expected. Other SU(2) subspaces do not exhibit spin squeezing.

\begin{figure}[t]
    \centering
    \includegraphics[width=\linewidth]{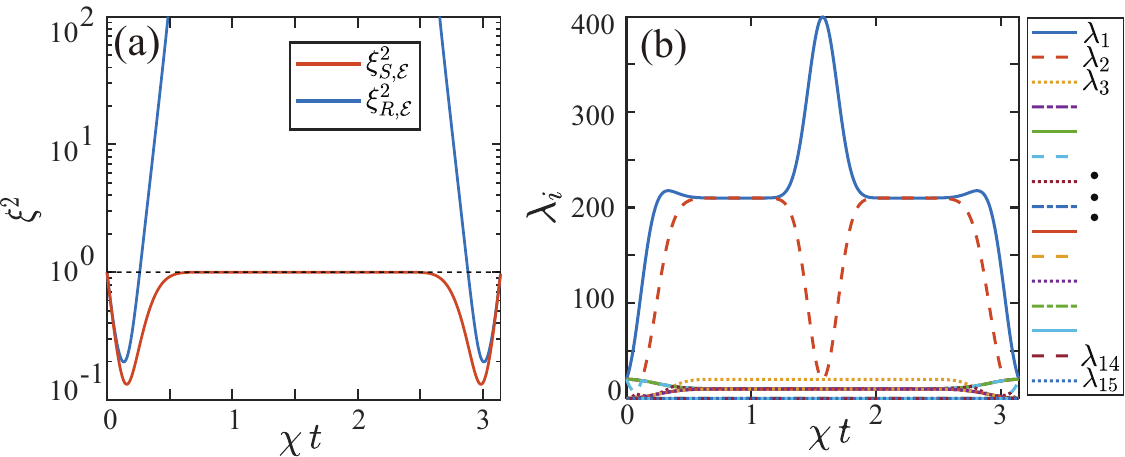}
    \caption{Dynamics of $\boldsymbol{\mathcal{E}}$-space coherent spin state for $\theta = 0$ and $N = 20$. (a) Evolution of the Kitagawa-Ueda squeezing parameter $\xi_{S,\mathcal{E}}^2$ and Wineland squeezing parameter $\xi_{R,\mathcal{E}}^2$. (b) Evolution of the QFI eigenvalue spectrum $\lambda_i$.}
    \label{fig:E_dynamics_theta0}
\end{figure}

Figure~\ref{fig:K_dynamics}(b) displays the eigenvalues' evolution of the QFI matrix $\mathbf{F}_Q$, where eigenvalues $\lambda_i$ are sorted in descending order ($\lambda_1 > \lambda_2 > \cdots > \lambda_{15}$). Two dominant eigenvalues ($\lambda_1$ and $\lambda_2$) are clearly visible, both significantly exceeding $N$, indicating the presence of two optimal metrological directions. Both optimal directions reside within the $\boldsymbol{\mathcal{K}}$ subspace, while other subspaces do not exhibit entanglement-enhanced precision during this dynamics.

From Figs.~\ref{fig:K_dynamics}(a) and (b), one observes that after reaching maximum squeezing, the system transitions into an over-squeezing regime where $\xi_{S,\mathcal{K}}^2 > 1$. The dominant QFI eigenvalue $\lambda_1$ keeps increasing during this over-squeezing period.
At the half-period ($\ t = T/2$), $\lambda_1$ reaches the maximum $N^2 = 400$, demonstrating Heisenberg quantum-enhanced sensitivity. Fig.~\ref{fig:K_dynamics}(c) shows the projection coefficients of its corresponding eigenvector $\mathbf{v}_1$ onto the 15 generator basis, where the colorbar represents the amplitude of projection coefficients. The optimal metrological direction undergoes a transition at $ t \approx 0.3 T$: initially residing in the $K_y$-$K_z$ plane, it switches to align purely along the $K_x$ direction for $0.3 T \lesssim  t<T/2$.

Figure~\ref{fig:K_dynamics}(d) demonstrates the scaling behavior of the maximum eigenvalue with particle number $N$ at two specific moments. At $ t = T/2$ (hollow circles), the scaling strictly follows the Heisenberg limit $N^2$, while at $ t = T/4 $ (square markers), the scaling behavior $0.63 N^{1.94}$ closely approaches the Heisenberg limit.

\subsection{Dynamics of $\boldsymbol{\mathcal{E}}$-Space Coherent Spin State}
\label{sec:e_dynamics}

We now focus on the dynamics of the $\boldsymbol{\mathcal{E}}$-space coherent spin state $|\text{CSS}\rangle_{\boldsymbol{\mathcal{E}}}$ defined in Eq.~(\ref{eq:CSS_E}), i.e., the initial state with particle occupations only in the $\hat{a}$ and $\hat{b}$ modes. Compared to the $\boldsymbol{\mathcal{K}}$-space CSS dynamics, the $\boldsymbol{\mathcal{E}}$-space case would exhibit richer behavior for a nonvanishing $\theta$, involving multiple SU(2) subspaces and non-periodic evolutions.

First, we examine the case $\theta = 0$. In this limit, one-axis-twisting spin squeezing emerges periodically in the $\boldsymbol{\mathcal{E}}$ subspace, while other subspaces exhibit no significant spin squeezing. Figs.~\ref{fig:E_dynamics_theta0}(a) and (b) show the temporal evolution of the spin squeezing parameters and QFI eigenvalue spectrum $\lambda_i$, respectively. The behavior closely resembles that observed in Figs.~\ref{fig:K_dynamics}(a) and (b) for the $\boldsymbol{\mathcal{K}}$-space case. However, unlike the $\boldsymbol{\mathcal{K}}$-space dynamics, the $\boldsymbol{\mathcal{E}}$-space one-axis twisting does not require finite $\theta$ and exhibits a fixed period $T = \pi/\chi$. This can be understood from the Hamiltonian in Eq.~(\ref{eq:hamiltonian_interaction_picture}): when $\theta = 0$, the {$\hat{N}_a \hat{N}_b$ term becomes $= -\frac{\chi}{4} (\hat{N}_a - \hat{N}_b)^2 = -\chi \hat{\mathcal{E}}_z^2$}, generating the characteristic one-axis twisting dynamics. Similarly, the projection of the maximum eigenvector $\mathbf{v}_1$ onto the SU(4) generators shows that the optimal metrological directions are concentrated in the $\boldsymbol{\mathcal{E}}$ subspace, with no contributions from other subspaces.

\begin{figure}[t]
    \centering
    \includegraphics[width=\linewidth]{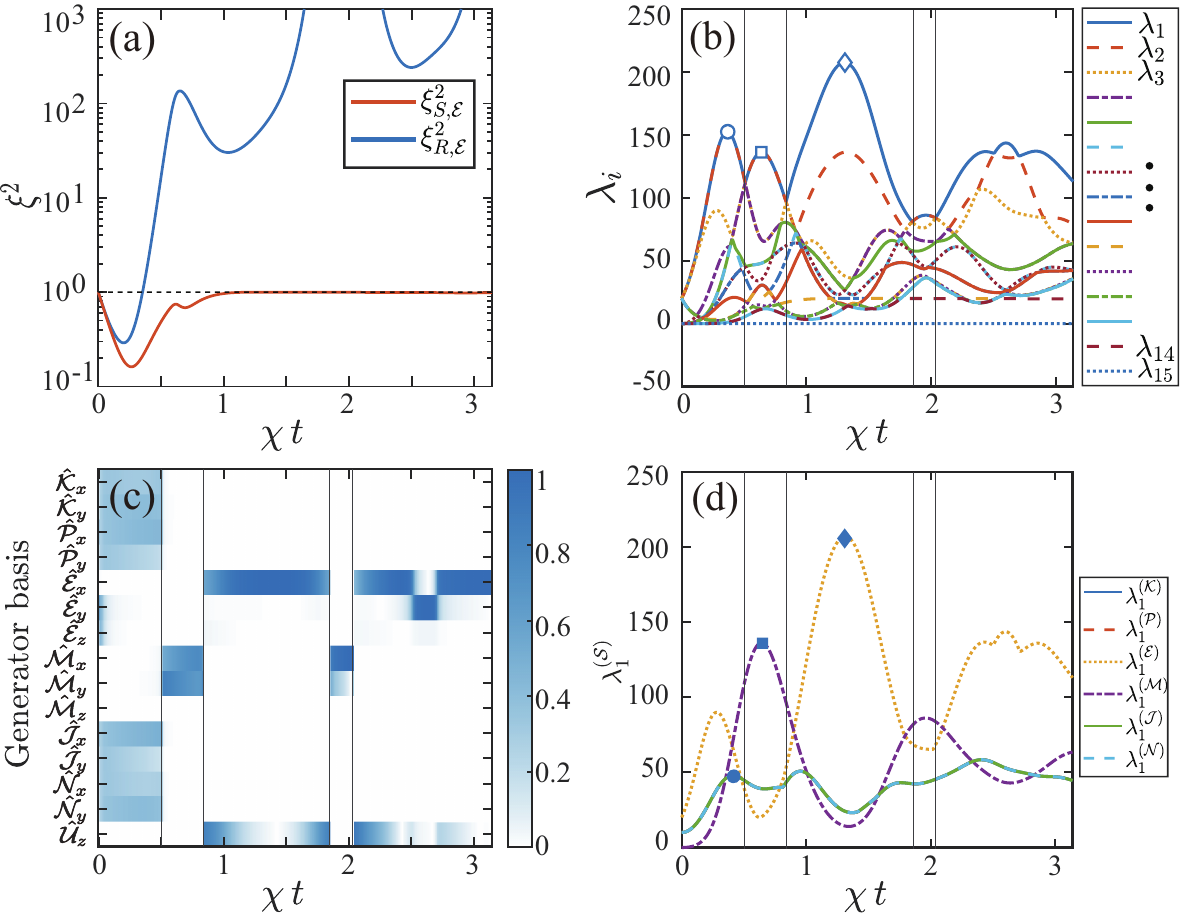}
    \caption{Dynamics of $\boldsymbol{\mathcal{E}}$-space coherent spin state for $\theta = \pi/8$ and $N = 20$. (a) Evolution of spin squeezing parameters. (b) QFI eigenvalue spectrum $\lambda_{i=\{1,\cdots,15\}}$ with multiple eigenvalues exceeding $N$. (c) Projection coefficients of the first eigenvector $\mathbf{v}_1$ onto the 15 generator basis. (d) Maximum eigenvalues $\lambda_{1}^{(\mathcal{S})}$ for SU(2) subspaces $\mathcal{S} \in \{\boldsymbol{\mathcal{K}}, \boldsymbol{\mathcal{P}}, \boldsymbol{\mathcal{E}}, \boldsymbol{\mathcal{M}}, \boldsymbol{\mathcal{J}}, \boldsymbol{\mathcal{N}}\}$. Vertical lines in panels (b), (c), and (d) delineate different dynamical stages.}
    \label{fig:E_dynamics_theta_pi8}
\end{figure}

The situation becomes significantly richer when $\theta \neq 0$. Figs.~\ref{fig:E_dynamics_theta_pi8}(a)-(c) respectively show the dynamics of the $\boldsymbol{\mathcal{E}}$-space squeezing parameters, QFI spectrum $\lambda_i$, and the maximum eigenvector $\mathbf{v}_1$ for $\theta = \pi/8$. As evident from the figures, both the spin squeezing parameters and QFI spectrum lose periodic structure. Spin squeezing appears only at short times, followed by a persistent over-squeezing regime.

The QFI spectrum $\lambda_i$ reveals numerous eigenvalues exceeding $N = 20$, indicating the presence of multiple quantum sensing capabilities. More importantly, as time goes, the QFI eigenvalues undergo crossovers, causing the optimal metrological direction $\mathbf{v}_1$ to switch between different subspaces, as illustrated in Fig.~\ref{fig:E_dynamics_theta_pi8}(c). In Figs.~\ref{fig:E_dynamics_theta_pi8}(b) and (c), we use vertical lines to divide the evolution time into distinct stages: In the first stage ($t \lesssim 0.51$), the optimal measurement directions reside in the $\boldsymbol{\mathcal{N}}$, $\boldsymbol{\mathcal{J}}$, $\boldsymbol{\mathcal{P}}$, and $\boldsymbol{\mathcal{K}}$ subspaces. Subsequently, for $0.51 \lesssim t \lesssim 0.84$, the optimal measurement direction shifts to the $\boldsymbol{\mathcal{M}}$ subspace. When $0.84 \lesssim t \lesssim 1.86$, the direction moves to the $\boldsymbol{\mathcal{E}}$ subspace, followed by a return to the $\boldsymbol{\mathcal{M}}$ subspace and then back to the $\boldsymbol{\mathcal{E}}$ subspace. This demonstrates that all SU(2) subspaces possess quantum metrological advantage.

\begin{figure}[t]
    \centering
    \includegraphics[width=\linewidth]{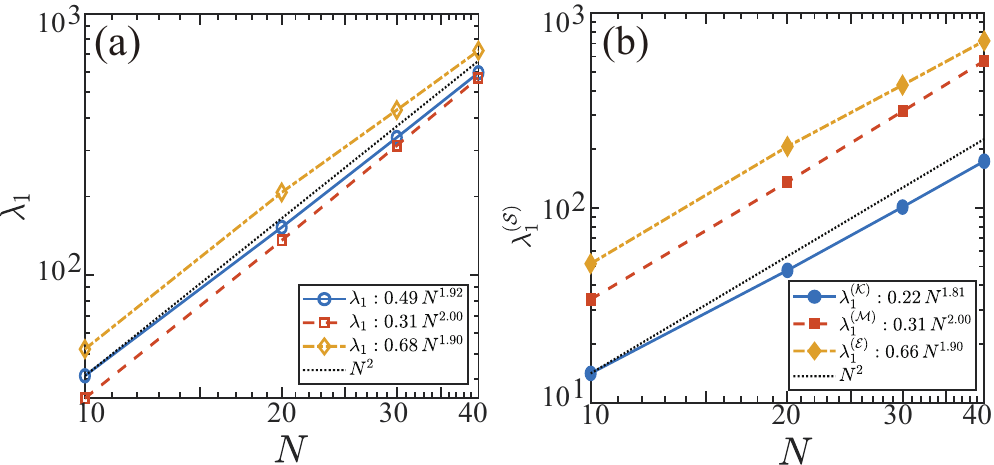}
    \caption{Scaling behavior of QFI eigenvalues $\lambda_1$ and $\lambda_1^{(\mathcal{S})}$ at selected moments with $\theta = \pi/8$. (a) Scaling of $\lambda_1$ at selected moments shown by hollow markers in Fig.~\ref{fig:E_dynamics_theta_pi8}(b). (b) Scaling of $\lambda_1^{(\mathcal{S})}$ at selected moments shown by solid markers in Fig.~\ref{fig:E_dynamics_theta_pi8}(c). The dashed black line indicates the Heisenberg scaling $N^2$.}
    \label{fig:E_scaling}
\end{figure}

To further clarify the QFI behavior in each subspace, we diagonalize the $3 \times 3$ QFI matrices $\mathbf{F}_Q^{({\mathcal{S}})}$ in the six SU(2) subspaces separately and present the maximum eigenvalues $\lambda_{1}^{(\mathcal{S})}$ in Fig.~\ref{fig:E_dynamics_theta_pi8}(d), with vertical lines consistent with panels (b) and (c). The curves $\lambda_{1}^{(\mathcal{S})}$ for the $\mathcal{S} \in \{\boldsymbol{\mathcal{K}}, \boldsymbol{\mathcal{P}}, \boldsymbol{\mathcal{J}}, \boldsymbol{\mathcal{N}}\}$ subspaces are overlapping. All subspaces exhibit maximum QFI eigenvalues exceeding $N = 20$. The $\boldsymbol{\mathcal{M}}$ subspace dominates the 2nd and 4th stages, while the $\boldsymbol{\mathcal{E}}$ subspace dominates the 3rd and final stages. In the first stage, although the eigenvalues of the $\lambda_{1}^{(\mathcal{S})}$ for $\mathcal{S} \in \{\boldsymbol{\mathcal{K}}, \boldsymbol{\mathcal{P}}, \boldsymbol{\mathcal{J}}, \boldsymbol{\mathcal{N}}\}$ are not individually maximal, the strong inter-coupling (correlations) among these four subspaces leads to the maximum eigenvalue $\lambda_1$ dominating the QFI spectrum after diagonalization, as shown in Fig.~\ref{fig:E_dynamics_theta_pi8}(b).

Next, we study the scaling properties of the optimal QFI eigenvalue on the particle number $N$ within various stages. Fig.~\ref{fig:E_scaling}(a) shows $\lambda_1$ versus $N$ at selected moments as indicated by hollow markers in Fig.~\ref{fig:E_dynamics_theta_pi8}(b), which correspond to the maximum values in the former three stages. 
Fig.~\ref{fig:E_scaling}(b) shows $\lambda_1^{(\mathcal{S})}$ versus $N$ at selected moments as indicated by solid markers in Fig.~\ref{fig:E_dynamics_theta_pi8}(d), which also correspond to the maximum values in subspaces. As demonstrated in Fig.~\ref{fig:E_scaling}(a), the spectrum of the complete QFI matrix $\mathbf{F}_Q$ approaches the Heisenberg limit at all three moments, with scaling factors being approximately $N^{1.92}$, $N^{2.0}$, and $N^{1.9}$, respectively. For the subspace QFI matrices $\mathbf{F}_Q^{(\mathcal{S})}$ shown in Fig.~\ref{fig:E_scaling}(b), the three moments exhibit scaling behaviors of $N^{1.81}$, $N^{2.0}$, and $N^{1.9}$, respectively. As can be inferred from Fig.~\ref{fig:E_dynamics_theta_pi8}(c), the 1st stage involves inter-correlations (coupling) among generators from different subspaces, while the subspace QFI matrices $\mathbf{F}_Q^{(\mathcal{S})}$ fail to capture these correlations, leading to the relatively reduced scaling factors, i.e., $1.81 < 1.92$.

\begin{figure}[t]
    \centering
    \includegraphics[width=\linewidth]{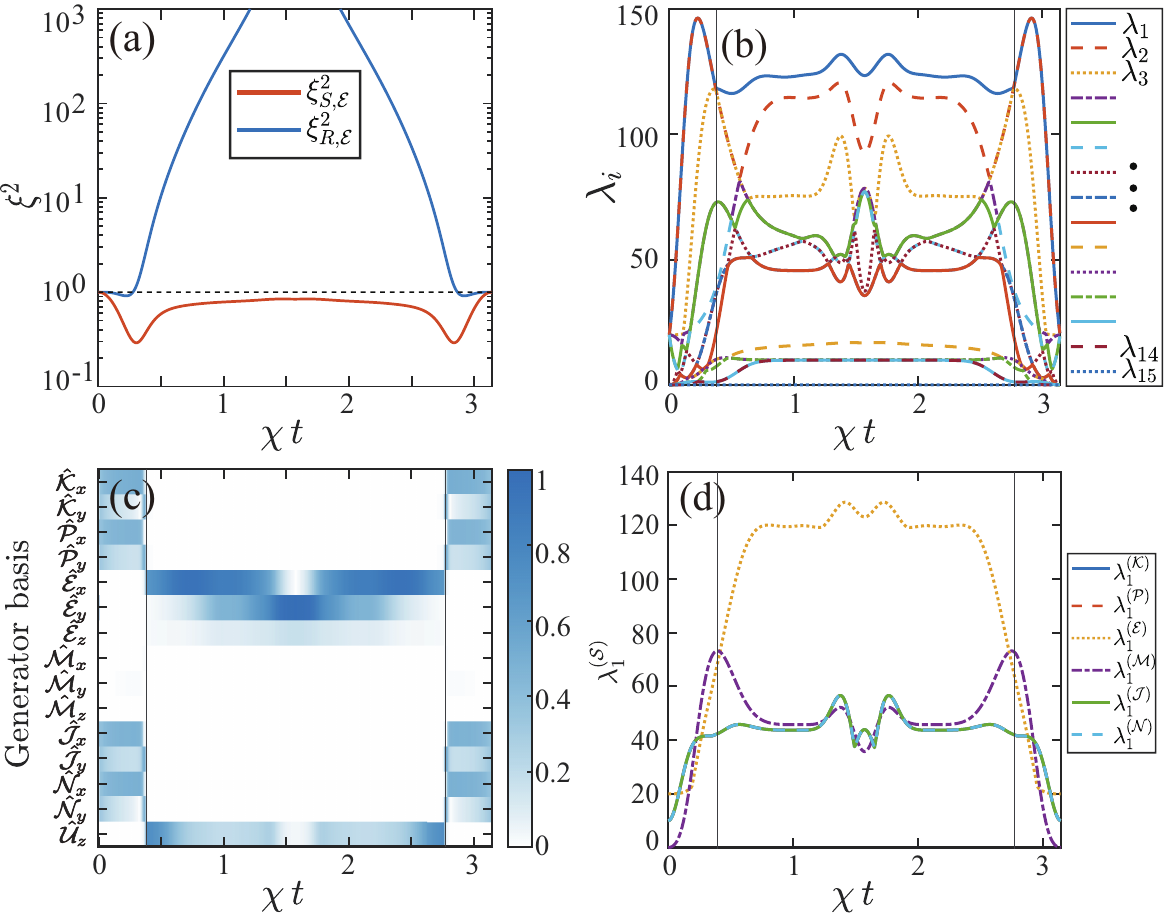}
    \caption{Dynamics of $\boldsymbol{\mathcal{E}}$-space coherent spin state for $\theta = \pi/4$ and $N = 20$. (a) Evolution of spin squeezing parameters. (b) Dynamics of QFI eigenvalue spectrum $\lambda_{i=\{1,\cdots,15\}}$. (c) Projection coefficients of $\mathbf{v}_1$ onto the 15 generator basis. (d) Maximum eigenvalues $\lambda_{1}^{(\mathcal{S})}$ for SU(2) subspaces.}
    \label{fig:E_dynamics_theta_pi4}
\end{figure}

After the detailed discussion on the case of $\theta = \pi/8$, we now turn to the case of $\theta = \pi/4$, with the results of squeezing parameters, QFI spectra $\lambda_i$, the projection amplitudes of $\mathbf{v}_1$, and $\lambda_1^{(\mathcal{S})}$ being displayed in Figs.~\ref{fig:E_dynamics_theta_pi4}(a)-(d), respectively. A prominent feature is that the dynamics again exhibit periodicity, with $T = \pi/\chi$. Similar to the case of $\theta = \pi/8$ (see Fig.~\ref{fig:E_dynamics_theta_pi8}), there are numerous eigenvalues $\lambda_i > 20$, and the maximum eigenvalue $\lambda_1$ is also able to switch between different subspaces. The difference is that the $\boldsymbol{\mathcal{M}}$ subspace makes no contribution to the maximum eigenvector $\mathbf{v}_1$ [projection coefficient $\sim 0$ in Fig.~\ref{fig:E_dynamics_theta_pi4}(c)]. However, the $\boldsymbol{\mathcal{M}}$ subspace still possesses metrologically useful entanglement, as shown in Fig.~\ref{fig:E_dynamics_theta_pi4}(d). It turns out that the $\boldsymbol{\mathcal{M}}$ subspace mainly contributes to the 5th eigenvector $\mathbf{v}_5$ (not shown). 

The phenomenon that all SU(2) subspaces exhibit quantum metrological advantage for $\theta > 0$ requires some additional explanation. This phenomenon does not exist for the dynamics with $|\text{CSS}\rangle_{\boldsymbol{\mathcal{K}}}$ as the initial state, as shown in the previous subsection (Sec.~\ref{sec:k_dynamics}). Here, the four-mode coupling terms in the effective Hamiltonian Eq.~(\ref{eq:hamiltonian_interaction_picture}) play a crucial role. For the dynamics with $|\text{CSS}\rangle_{\boldsymbol{\mathcal{E}}}$ [Eq.~(\ref{eq:CSS_E})] as the initial state, although the initial state only has particle occupations in the $\hat{a}$ and $\hat{b}$ modes (within the $\boldsymbol{\mathcal{E}}$ subspace), particles can transfer to the $\hat{c}$ and $\hat{d}$ modes through the four-mode scatterings $(\hat{a}^{\dagger} \hat{b}^{\dagger} \hat{c} \hat{d} + \text{h.c.})$, thereby enabling all subspaces to be metrologically valuable. However, for the dynamics initialized with $|\text{CSS}\rangle_{\boldsymbol{\mathcal{K}}}$ [Eq.~(\ref{eq:CSS_K})], there is no corresponding four-mode scattering channel to transfer particles to the other two modes ($\hat{a}$ and $\hat{c}$), and hence the metrological advantage can only be manifested in the $\boldsymbol{\mathcal{K}}$ subspace defined by the $\hat{b}$ and $\hat{d}$ modes. This explains why, for any value of $\theta$, the dynamics of $|\text{CSS}\rangle_{\boldsymbol{\mathcal{K}}}$ only exhibits metrologically useful entangled states only in the $\boldsymbol{\mathcal{K}}$ space.

\begin{figure}[t]
    \centering
    \includegraphics[width=\linewidth]{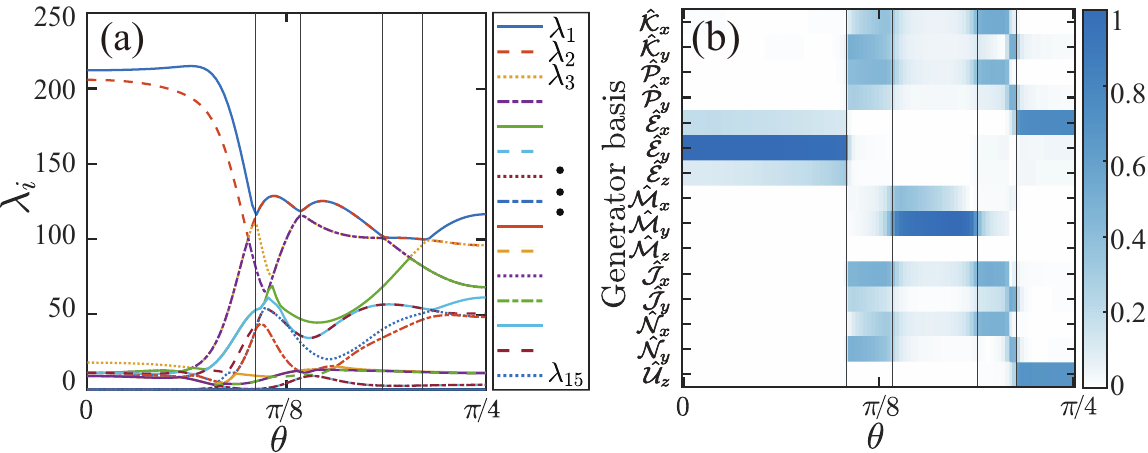}
    \caption{Parameter dependence of $\boldsymbol{\mathcal{E}}$-space coherent spin state dynamics at fixed evolution time $ t = 0.48$ with $N=20$. (a) QFI eigenvalue spectrum $\lambda_{i=\{1,\cdots,15\}}$ as a function of dressing angle $\theta$. (b) Projection coefficients of the first eigenvector $\mathbf{v}_1$ onto the 15 generator basis as a function of $\theta$. Vertical lines mark different dynamical stages.}
    \label{fig:E_theta_dependence}
\end{figure}

In the above, we have analyzed the dynamics of QFI for given values of $\theta$. Finally, let us briefly discuss influence of $\theta$ on the final state with a given total evolution time $t$. The results demonstrate that by properly choosing the evolution time $t$, one can also observe quantum sensing capabilities distributed across various subspaces. For example, by fixing $ t = 0.48$, we show in Fig.~\ref{fig:E_theta_dependence} the QFI full spectra $\lambda_i$ and the projection coefficients of the maximum eigenvector $\mathbf{v}_1$ as functions of $\theta$. From panel (a), one can observe a clear transition at $\theta = 0.33$, where the QFI spectrum changes from being dominated by two eigenvalues to being dominated by multiple eigenvalues. This again confirms that the four-mode scattering induced by a non-vanishing $\theta$ is crucial for inducing and distributing metrological resources across multiple subspaces. Additionally, panel (b) shows that the optimal metrological direction can switch between different SU(2) subspaces [similar to Fig.~\ref{fig:E_dynamics_theta_pi8}(c)], with the switching points again marked by vertical lines.

\section{Discussion and Conclusion}
\label{sec:conclusion}

The experimental measurement of the SU(4) generators deserves consideration. Measurements in any SU(2) subspace are experimentally feasible, and all SU(4) generators are simply linear combinations of the subspace generators. For any subspace, the $z$-direction operators $\hat{\mathcal{S}}_z$ is a particle number operator, which can be directly measured through standard detection techniques. Time-of-flight imaging \cite{Chu1986, Yavin2002} enables the distinction of particles with different momenta, while Stern-Gerlach imaging can resolve different spin states. 
Additionally, by introducing external fields, one can independently rotate spin and momentum degrees of freedom, facilitating arbitrary-direction measurements within subspaces, i.e., spin-noise tomography \cite{Ariano2003,Riedel2010,Gross2010}.
Specifically, since the up and down spin states generally correspond to two hyperfine ground states of atoms, spin control can be achieved by introducing a radio-frequency pulse at frequency $E_+(k_0) - E_-(k_0) \sim \text{kHz}$, which leads to $\hat{\boldsymbol{\Psi}}^\dagger \hat{\sigma}_x\hat{\boldsymbol{\Psi}} = \cos(2\theta)(\hat{a}^\dagger \hat{d} - \hat{b}^\dagger \hat{c} + \text{h.c.})$ that purely couples the upper- and lower-band modes without affecting the momentum distribution. 
For momentum control, the auxiliary optical lattice approach used in Refs.~\cite{engles1,engles2,chenSpinSqueezingSpinorbitcoupled2020} can be employed, where an external potential $\cos^2(k_0 x) \sim V_0(e^{2ik_0x} + e^{-2ik_0x})/4$ is added to the BEC, effectively coupling the states $\pm |k_0\rangle$.

An additional point that requires clarification is that, although we have presented the results of $\theta = \pi/4$ in the above, this value is actually not experimentally achievable. When $\theta = \pi/4$ (such that $\Omega = 4E_r$), the lower single-particle band $E_-$ is no longer in the double-well structure. Both $\pm k_0$ approach $k_0 = 0$, and consequently the four-mode model becomes ill-defined. However, from a theoretical perspective, $\theta = \pi/4$ represents a meaningful reference point that helps us capture the physics near $\theta = \pi/4$.

To conclude, this work demonstrates that spin-orbit coupling provides a controllable degree of freedom for spinor BECs to realize a four-mode model that can be described by an $\mathfrak{su}(4)$ algebraic containing six SU(2) subspaces. Compared to traditional two-mode models, the four-mode model potentially demonstrates richer metrologically useful quantum resources in both the large SU(4) and individual SU(2) subspaces, with QFI spectra approaching the Heisenberg scaling limit. The spin-orbit-coupling parameter $\theta$ (or equivalantly the Raman Rabi frequency) plays a crucial role: it provides four-mode coupling terms that enable metrologically useful entanglement to emerge in all SU(2) subspaces, even when the initial state has particle distributions in only two modes. By appropriately controlling $\theta$ or the evolution time, one can engineer the maximal QFI and the optimize measurement directions across different subspaces.

\begin{acknowledgments}
L. C. acknowledges support from the NSF of China (Grant No. 12174236) and from the fund for the Shanxi 1331 Project. 
\end{acknowledgments}

\end{document}